\documentclass[a4paper,11pt]{revtex4}
\usepackage{epsfig}

\newcommand{\be}{\begin{equation}}
\newcommand{\ee}{\end{equation}}
\newcommand{\bea}{\begin{eqnarray}}
\newcommand{\eea}{\end{eqnarray}}
\newcommand{\bean}{\begin{eqnarray*}}
\newcommand{\eean}{\end{eqnarray*}}

\newcommand{\gapproxeq}{\lower
.7ex\hbox{$\;\stackrel{\textstyle >}{\sim}\;$}}
\newcommand{\lapproxeq}{\lower
.7ex\hbox{$\;\stackrel{\textstyle <}{\sim}\;$}}

\begin{document}

\title{\bf Locality of quark-hadron duality and deviations
from quark counting rules above resonance region}

\author{Qiang Zhao$^1$\footnote{e-mail: Qiang.Zhao@surrey.ac.uk}
and Frank E. Close$^2$\footnote{e-mail: F.Close1@physics.ox.ac.uk}
}
\affiliation{1) Department of Physics, 
University of Surrey, Guildford, GU2 7XH, United Kingdom}
\affiliation{2) Department of Theoretical Physics,
University of Oxford, \\
Keble Rd., Oxford, OX1 3NP, United Kingdom}

\date{\today}
  
\begin{abstract}

We show how deviations 
from the dimensional scaling laws
for exclusive processes may be related to a breakdown in the locality of 
quark-hadron duality. The essential principles are illustrated
in a pedagogic model of 
a composite system with two spinless charged constituents, 
for which a dual picture for the low-energy resonance phenomena 
and high-energy scaling behavior can be established.
We introduce the concept of ``restricted locality" of
quark-hadron duality and show how this 
results in deviations from the pQCD quark counting rules
above the resonance region. In particular it can be a possible 
source for oscillations about the smooth quark counting
rule, as seen e.g. in 
 the 90-degree differential cross sections
for $\gamma p\to \pi^+ n$.

\end{abstract}

\maketitle  



The dimensional scaling laws~\cite{brodsky-farrar-75,matveev} 
have had considerable success
in high energy exclusive scattering, where a valence-like minimal 
number of quarks is probed. Within this model,
dominance of short distance pQCD phenomena
could be expected at quite low energies, for example, immediately above
the traditional resonance regions.  
On the other hand, the empirical status of 
the helicity conservation selection rules
of Lepage and Brodsky~\cite{l-b-80} 
at these low energies is unclear.
The assumption of short distance 
 dominance for such exclusive processes has also been questioned theoretically, e.g. 
Ref.~\cite{isgur-llewelyn-smith-84} argues that 
 nonperturbative processes could still be important
 in some kinematic regions even at 
high energies. 
In summary: the transition between perturbative and 
strong interaction regimes of QCD is obscure.

In experiment, deviations from quark counting rules 
have been found in exclusive reactions~\cite{exclusive-exp}. 
In particular,
the energy dependence of data at $\theta_{c.m.}=90^\circ$
oscillates
around the value predicted by the quark counting rules.
Explanations include 
the opening of new flavor channels~\cite{b-de-88}, 
interference between pQCD 
and long-distance components~\cite{b-c-l}, 
and the essential message is that a sizeable long-distance 
interaction component
cannot be neglected~\cite{miller-02,belitsky}. 
Also, pQCD color transparency ~\cite{r-p-82-88,ksjr-00}
predicts interesting phenomena.

In this Letter, we show how recent 
ideas~\cite{IJMV,melnitchouk,ClIs01,JV,close-zhao} 
on quark-hadron duality~\cite{bloom,duality,jlab1}
can give novel insights into the derivation of counting rules, 
at least in non-diffractive processes.
We shall explore
the intermediate high energy region,
through which we shall relate resonance excitations at low energies
to parton phenomena at high energies.
This leads to a smooth counting-rule type of behavior at 
$\theta_{c.m.}=90^\circ$
modulated by oscillations, as observed.

To illustrate the essential idea,
the recently developed model of
two-body spinless constituents~\cite{IJMV,ClIs01,JV,close-zhao}
serves as the simplest example for the realization of duality.
The general form for the transition amplitude 
for $\gamma({\bf k}) \Psi_0\to \Psi_N\to \Psi_0 \gamma ({\bf q})$ can be
expressed as
\bea
\label{non-forward-trans}
M&=& 
\sum_N\langle\Psi_0({\bf P}_f,{\bf r})|[e_1e^{-i{\bf q}\cdot{\bf r}/2}
+e_2e^{i{\bf q}\cdot{\bf r}/2}]|\Psi_N\rangle
\nonumber\\
&&\times\langle \Psi_N|  [e_1e^{i{\bf k}\cdot{\bf r}/2}
+e_2e^{-i{\bf k}\cdot{\bf r}/2}]|\Psi_0({\bf P}_i, {\bf r})\rangle \ ,
\eea
where $\Psi_N$
is the harmonic oscillator wave function with the main quantum number $N$.
We shall abstract some general features from the above model.
In ~\cite{ClIs01,close-zhao}, it was shown that
the sum over the resonances can be 
related to the scaling behavior as a result of destructive interferences
among different resonances 
in the coherent terms ($\sim e_1 e_2$) at small $|t|$.
At high energies the destructive interference is rather local, due to
the high density of overlapping resonances.
On the other hand, at low energies 
only a few resonances contribute, 
there is marked breaking of the
mass degeneracy of states with the same $N$, and orbital angular
momentum
$L$-dependence of the resonances plays an important role.

The imaginary part of Compton scattering 
as a sum over the intermediate ``resonance" states 
in Eq.~(\ref{non-forward-trans}) accesses the structure function of 
``nucleon" $\Psi_0$. As discussed in \cite{close-zhao}, 
taking the $z$-axis along the incoming photon momentum direction,
and explicitly including the $L$-dependence 
in the energy spectrum, 
we can express the transition amplitude as
\bea
\label{trans-01}
M&=&\sum_{N=0}^{\infty}  
\left[(e_1^2+e_2^2)\frac{1}{N !}
\left(\frac{{\bf k}\cdot{\bf q}}{2\beta^2}\right)^N
\right.\nonumber\\
& + & 2e_1e_2\frac{1}{N !}
\left.\left(-\frac{{\bf k}\cdot{\bf q}}{2\beta^2}\right)^N
\right]
C_{N} e^{-({\bf k}^2+{\bf q}^2)/4\beta^2}\nonumber\\
&=&
\sum_{N=0}^{\infty} \sum_{L=0(1)}^N 
[(e_1^2+e_2^2)d_{00}^L(\theta) +2e_1e_2d_{00}^L(\pi-\theta)]\nonumber\\
&\times &
C_{NL} {\cal F}_{0N}^{(L)}({\bf q})  
{\cal F}_{N0}^{(L)}({\bf k})   \ ,
\eea
where ${\cal F}_{N0}^{(L)}({\bf k})$ denotes the transition 
(with momentum ${\bf k}$) to an excited state with 
quantum number $(N,L)$, while ${\cal F}_{0N}^{(L)}({\bf q})$
denotes the decay back to the ground state.
In Compton scattering, 
these two transition amplitudes are connected by a Wigner rotation 
function $d_{00}^L(\theta)$, where $\theta$ is 
the relative angle between the incoming and outgoing photon 
momenta ${\bf k}$ and ${\bf q}$
in the photon-target c.m. system.
The factorized $L$-dependence (i.e. violated $C_N$)
is denoted by a coefficient $C_{NL}$
which is essentially related to the non-degenerate mass positions
of excited resonant states, and is the same for 
the coherent and incoherent terms in
the two-body system.

At this stage, it is not important to concern ourselves with details of the 
$L$-dependent factor $C_{NL}$. 
First note that in this simple model~\cite{ClIs01,close-zhao}
all terms of $L=odd$ for a given $N$
are proportional to $\cos\theta$, and hence vanish at $\theta=90^\circ$;
thus
we need consider only the parity-even 
states at $\theta=90^\circ$, 
i.e. $N=0$, 2, 4 ... with $L=N$, $N-2$, $\cdots$ 0 for a given even $N$. 
The scattering amplitude at $90^\circ$ can then be expressed as
\bea
\label{trans-even}
&&(M_{N=0}+M_{N=2}+M_{N=4}+\cdots)_{\theta=90^\circ}\nonumber\\
&=&e_0^2\left[ C_{00}\left(\frac{kq}{2\beta^2}\right)^0
+\frac{1}{2!}\frac 13(-C_{22}+C_{20})\left(\frac{kq}{2\beta^2}\right)^2
\right. \nonumber\\
&&\left.+\frac{1}{4!}\frac{1}{35}(3C_{44}-10C_{42}+7C_{40})
\left(\frac{kq}{2\beta^2}\right)^4 
+\cdots \right]\nonumber\\
&&\times e^{-({\bf k}^2 + {\bf q}^2)/4\beta^2} \ ,
\eea
where $e_0=e_1+e_2$ is the total charge of ``nucleon" $\Psi_0$.

Several points thus can be learned:

i) At high energies where the state degeneracy limit can be applied, 
all the terms with $N\ne 0$ and $L=0,\cdots, N$ in
Eq.~(\ref{trans-even}) would vanish due to the destructive cancellation.
Only the $C_{00}$ term survives:
\be
\label{degenerate}
M= e_0^2C_{00} e^{-\frac{({\bf k}-{\bf q})^2}{4\beta^2}}\Big|_{\theta=90^\circ}
= e_0^2 C_{00} R(t)\Big|_{\theta=90^\circ} \ ,
\ee
where $R(t)$ is recognised as the elastic form factor for the Compton 
scattering~\cite{close-zhao} or more generally the quark counting rule
predicted 
scaling factor~\cite{brodsky-farrar-75,matveev}. So we see that the smooth
behaviour driven by the elastic form factor, which is the essence of the
counting 
rules, effectively arises from the {\it s}-channel sum combined with 
the destructive interferences among resonances.

ii) Concerning the $L$-degeneracy breaking effect for any given $N$,
each term of $(N,L)$ corresponds to the excitation 
of an intermediate state with given $N$ and $L$. 
The factor $C_{NL}$, which is essentially related to the 
mass position of each state, should be different for the individual
states. 
This leads to oscillations around the simple result of Eq.~(\ref{degenerate}),
due to different partial waves not cancelling locally.
We shall refer to this as ``restricted locality".

Certainly, the simple model can only illustrate 
such a deviation in a pedagogic way. However, 
a similar phenomenon may have existed in physical processes due to 
the restricted locality of duality above the prominent 
resonance region. Notice that deviations from 
quark counting rules are indeed found experimentally in 
certain exclusive reactions~\cite{experiment-data,gao}: for example,
the 90$^\circ$ differential cross sections of 
$\gamma p\to \pi^+ n$ at $W\sim 3$ GeV exhibit oscillations around
the scaling curves predicted by the counting rules.
We will show how the restricted locality of duality is naturally a
source 
of such oscillations.

To generalize the above to the physical exclusive processes,
we adopt effective Lagrangians
for the constituent-quark-meson and quark-photon couplings, while 
treating the mesons as elementary particles, as e.g., the effective 
theory proposed by Manohar and Georgi~\cite{manohar} and extended 
to pseudoscalar meson production in Refs.~\cite{Li-94,li-97,zhao-pion}.
Briefly, the introduction of an effective Lagrangian for quark-meson
couplings
highlights the quark correlations
in the exclusive processes (including the Compton scattering). 
We can thus arrive at a general expression for the transition amplitudes
for the {\it s}- and {\it u}-channels, i.e. the direct 
and the virtual resonance excitations:
\begin{eqnarray}
\label{s-deg}
M^{s+u}_{fi}&=& e^{-({\bf k}^2+{\bf
q}^2)/6\alpha^2} \nonumber\\
&\times & 
\left\{ \sum_{n=0}^{\infty}({\cal O}^{cc}_d + (-\frac 12)^n {\cal O}^{cc}_c)
\frac{1}{n!}
\left(\frac{{\bf k}\cdot{\bf q}}{3\alpha^2}\right)^{n}\right.\nonumber\\
&+&\sum_{n=1}^{\infty} ({\cal O}^{ci}_d + (-\frac 12)^n {\cal O}^{ci}_c)
\frac{1}{(n-1)!}
\left(\frac{{\bf k}\cdot{\bf q}}{3\alpha^2}\right)^{n-1}\nonumber\\
&+&\left.\sum_{n=2}^{\infty} ({\cal O}^{ii}_d + (-\frac 12)^n {\cal O}^{ii}_c)
\frac{1}{(n-2)!}
\left(\frac{{\bf k}\cdot{\bf q}}{3\alpha^2}\right)^{n-2} \right\}\ ,
\end{eqnarray}
where the multiplets are degenerate in $n$.
The spin structures, charge and isospin operators have been 
subsumed in the symbol ${\cal O}$. 
Terms proportional to $({\bf k}\cdot{\bf q}/3\alpha^2)^n$
denote correlations of c.m. - c.m. motions (superscript $cc$), 
while $({\bf k}\cdot{\bf q}/3\alpha^2)^{n-1}$
and $({\bf k}\cdot{\bf q}/3\alpha^2)^{n-2}$ denote the 
c.m. - internal ($ci$) or internal - internal correlations ($ii$), respectively.
The subscript ``d" (``c") denotes the 
direct (coherent) process 
that the photon and meson couple to the same (different) quarks
in the transition. The coherent process
is suppressed by a factor of $(-1/2)^n$ in comparison with 
the direct one for higher excited states.
Note that the conventional Born terms will contribute
to  different parts: the nucleon pole terms included in the 
{\it s}- and {\it u}-channel, and the possible contact term and 
{\it t}-channel charged meson exchange 
included as part of the background terms
due to gauge invariance. Both terms can be expressed 
as 
\be
M^{c+t}_{fi}={\cal O}^{c+t} e^{-({\bf k}-{\bf q})^2/6\alpha^2} \ .
\ee
This expression is similar to 
that discussed in Ref.~\cite{close-zhao}
 for the simple 
two spinless constituent system. The detail of the 
exclusive process (i.e. the detail of the spin operators)
does not prevent us from recognising certain general aspects
of the amplitude, in particular the effects that a
restricted locality of duality has
in the interplay between the resonance and partonic phenomena.

In the low energy regime, the degeneracy in $n$ must 
break. In the SU(6)$\otimes$O(3) symmetry limit, 
for a given $n$ ($\le 2$), multiplets of 
$L$- and $S$-dependent resonances can be separated in this model.
As studied in Ref.~\cite{zhao-pion},
quantitatively the calculations were in agreement
with experimental data up to $E_\gamma\approx$ 500 MeV.

The dominant term comes from the correlation of the c.m.- c.m. 
motions at the two vertices ($n=0, 1, \cdots$), while 
terms involving the c.m.- internal motion correlation, or 
internal - internal motion correlation will be suppressed. 
For example, for $n=0$ only the terms involving the c.m. - c.m. 
correlation contribute. These correlations are essentially
the demonstration of the internal degrees of freedom 
of the nucleon system. 

In the high energy limit where the degeneracy achieves, 
the leading term can be expressed
compactly as follows:
\begin{equation}
M^{s+u}_{fi}=
({\cal O}^{cc}_d + {\cal O}^{cc}_c e^{-{\bf k}\cdot{\bf q}/2\alpha^2})
e^{-({\bf k}-{\bf q})^2/6\alpha^2} \ ,
\end{equation}
where similar to Refs.~\cite{IJMV,ClIs01,JV,close-zhao} the scaling 
behavior can be realized at small $|t|$
due to the suppression of $e^{-{\bf k}\cdot{\bf q}/2\alpha^2}$
on the coherent term.
At $\theta=90^\circ$, we have
\begin{equation}
\label{s-asymp}
M^{s+u}_{fi} =  ({\cal O}^{cc}_d + {\cal O}^{cc}_c)
e^{-({\bf k}-{\bf q})^2/6\alpha^2}\Big|_{\theta=90^\circ} \ ,
\end{equation}
where both direct and coherent
process contribute and
operators ${\cal O}^{cc}_d$ and ${\cal O}^{cc}_c$ now
are independent of $n$. 
We conjecture that 
a similar factorization for the exclusive process 
may be more general than this
nonrelativistic pictures, as suggested by the pedagogic model~\cite{close-zhao}.
The form of Eq.~(\ref{s-asymp}) 
then represents the realization of duality, in particular, the emergence 
of the empirical quark counting rules after the sum over degenerate
resonances
at high energies.
The exponent factor $e^{-({\bf k}-{\bf q})^2/6\alpha^2}$ is thus 
regarded as the ``typical" scaling law factor.

This model has interesting implications
for kinematics just above the resonance region. Here, where 
the resonances of $n\ge 3$ are not degenerate, 
we expect to see effects of interference 
among non-local resonances, e.g.
states of $n=3$ and $n=4$, 
giving deviations from quark counting rules at
90$^\circ$. 

This conjecture seems likely to be realized given 
the evidence for higher excited states ($n>2$)~\cite{pdg2002},
where local degeneracy  
has not yet been reached at energies of a few GeVs. 
Therefore, the degeneracy breaking will cause 
deviations from the smooth counting rules
due to the different $L$-dependence in the resonance configurations.

\begin{figure}
\begin{center}
\epsfig{file=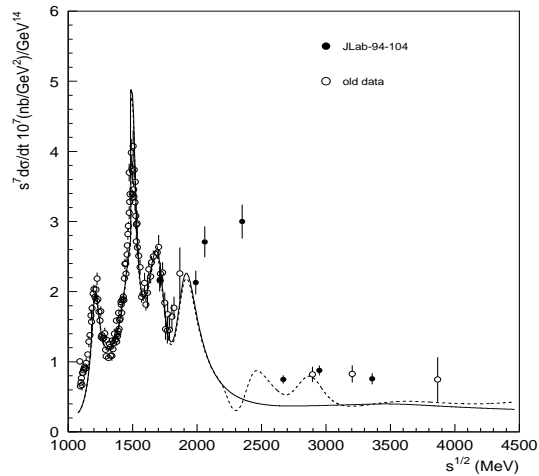, width=8cm,height=7.cm}
\caption{Energy dependence of the differential cross section
for $\pi^+$ photoproduction at $\theta=90^\circ$. The solid curve
denotes degeneracy breaking for $n\le 2$, while the dotted for $n\le 4$.
The empty circles are old data from Ref.~\cite{experiment-data},
and the full dottes are new data from JLab~\cite{gao}.
}
\protect\label{fig:(1)}
\end{center}
\end{figure}

To estimate the 90$^\circ$ deviation in $\gamma p\to \pi^+ n$,
we introduce
the mass-degeneracy breaking ($L$-dependence) 
into the $n=3$ and 4 terms in Eq.~(\ref{s-deg}).
For instance,
for $n=3$, we assume that 
the $L$-dependent multiplets are still proportional to 
${\bf k}\cdot{\bf q}$, thus vanish
 at $\theta=90^\circ$ as shown by Eq.~(\ref{trans-01}).
For $n=4$, one would have $P$, $F$ and $H$ partial waves.
The non-degeneracy then gives non-vanishing terms in the 90$^\circ$ cross
section.
Above the third resonance region, the quark model form factor is not 
an accurate representation of data. 
For $s\lesssim$ 20 GeV$^2$ 
at 90$^\circ$,
$R(t)\approx (0.22-t)/(0.025-t)$. So, empirically we make contact with
the counting rule by replacing $R(t)$ of Eq.~(\ref{s-deg}) by
\be
\frac{1}{(1-t/0.7)^2}\left(\frac{0.025-t}{0.22-t}\right)R(t).
\ee
This applies if the symmetry limit were true for all $n$.
Taking account of the non-degeneracy for $n\le 2$ gives the solid curve
in Fig.~\ref{fig:(1)}, which includes prominent well known resonances.
Including non-degeneracy for $n\le 4$~\cite{pdg2002} gives the dotted curve 
in Fig.~\ref{fig:(1)}. This shows how breaking the restricted locality
of duality produces sizeable oscillations that persist to a few GeVs of $W$,
but with reduced amplitudes as energy increases.
Such a result, even though very qualitative 
(we do not try to fit the data above the third 
resonance region~\cite{experiment-data}), 
suggests that non-degenerate higher resonances cause the 
deviations from the quark counting rules
above $W\approx 2.5$ GeV in meson photoproduction.
As a consequence, 
the $Q^2$-dependence of the oscillations could be most 
illuminating. For example, if a subset of resonances is relatively
suppressed at large $Q^2$ (as proposed in Ref.~\cite{CM}),
there will be significant shifts in the oscillations, 
both in position and magnitude, rather than the relatively stable 
changes predicted by color transparency.
In particular, and in contrast to other phenomenologies, 
the deviation pattern produced by the resonance degeneracy breaking
need have no simple periodicity.
The experimental data 
can thus distinguish this mechanism from others.

Additionally, if the main splitting mechanism 
were due to the partial wave dependence, 
at $\theta=90^\circ$ destructive interference
at high energies should occur within states of a given $n$, i.e. with the 
same parity. 
Consequently, parity-even and parity-odd states 
could be isolated at $\theta=90^\circ$ since the degenerate
terms are proportional to $({\bf k}\cdot{\bf q}/3\alpha^2)^n$.

To summarize: we have discussed the relation between resonance 
phenomena and the dimensional scaling laws based on the
quark-hadron duality picture 
at $2\lesssim \sqrt{s} \lesssim 3.5$ GeV. 
In contrast to previous 
models for the deviations from quark counting rules,
here we proposed that non-perturbative resonance excitations
are an important source for 
such deviations. At specific kinematics,
e.g. $\theta=90^\circ$, the oscillatory deviations could be
dominantly produced by resonance excitations with ``restricted 
locality". 
This argument is general for photon induced two-body 
reactions on the nucleon, and so we expect that such a restricted 
locality of duality occurs in vector meson photoproduction
at a few GeVs as well. The existence of higher excited resonances and
the recent experiment 
at JLab~\cite{clas-phi-00,clas-rho0-02,jlab-94,jlab-02}
suggest that the large angle cross sections 
for the photoproduction of the $\omega$, $\rho^0$ and $\phi$ 
at a few GeVs
are still dominated by  {\it s}- and {\it u}-channel 
nucleon exchanges~\cite{laget-00,zhao-phi-01}.
Although the formulation is nonrelativistic, 
we find it has been valuable to gain insights into 
the regime between the traditional resonance and partonic regions.
We also suggest a non-trivial $Q^2$ dependence for such oscillations.

This work is supported, in part, by grants from 
the U.K. Engineering and Physical
Sciences Research Council (Grant No. GR/R78633/01),
and the Particle Physics and
Astronomy Research Council.
We thank H. Gao for useful discussions and providing experimental data.



%

\end{document}